\documentclass[journal]{IEEEtran}
\usepackage{amsmath,graphicx}
\usepackage{epstopdf}
\usepackage{color}
\usepackage{wrapfig}
\usepackage{lipsum}
\usepackage{float}
\usepackage{colortbl,booktabs}
\usepackage{mathtools}
\usepackage[compress]{cite}
\definecolor{myRed}{rgb}{0.8, 0.2, 0.2}
\definecolor{myYellow}{rgb}{0.2,0.2,0.8}

\newcommand{\figref}[1]{Fig.~\ref{#1}}

\newcommand{\eqnref}[1]{(\ref{#1})}

\usepackage{amssymb}

\usepackage{multirow}
\usepackage{float}
 \usepackage{bm}
\usepackage{subfigure}
\usepackage{multicol}
\usepackage{algorithmic}
\usepackage{algorithm}
 \usepackage{mathrsfs}
\usepackage{bm}
\usepackage{amsthm}
\theoremstyle{definition}

\newtheorem{prop}{Proposition}
\theoremstyle{Proposition}
\theoremstyle{remark}
\newtheorem{remark}{Remark}

\theoremstyle{Corollary}
\usepackage{cite}
\usepackage{stfloats}%
\usepackage{lettrine}
\usepackage[margin=0.493in]{geometry}
\ifCLASSINFOpdf
\else
\fi
\begin{document}

\title{\huge Splitting Receiver with Joint Envelope and Coherent Detection}

%\author{Yanyan Wang, Member, Wanchun Liu, Member, Xiangyun Zhou, Senior Member}
\author{Yanyan Wang, \emph{Member, IEEE}, Wanchun Liu, \emph{Member, IEEE}, Xiangyun Zhou, \emph{Senior Member, IEEE}\vspace{-0.65cm}}
\maketitle
\IEEEpeerreviewmaketitle

\begin{abstract}
\let\thefootnote\relax\footnote{
	Y. Wang is with School of Information Science and Technology, Southwest Jiaotong University, Chengdu, China  (email: yanyanwang@swjtu.edu.cn).
	
W. Liu is with School of Electrical and Information Engineering, The University of
	Sydney, Australia (email: wanchun.liu@sydney.edu.au).
	
X. Zhou is with School of Engineering, The Australian National University, Australia  (email: xiangyun.zhou@anu.edu.au).	
}
This letter proposes a new splitting receiver design with joint envelope detection (ED) and coherent detection (CD). To characterize its fundamental performance limit, we conduct high signal-to-noise ratio (SNR) analysis on the proposed ED-CD splitting receiver and obtain closed-form approximations of both the achievable mutual information and the optimal splitting ratio (i.e., a key design parameter of the receiver). Our numerical results show that these high SNR approximations are accurate over a wide range of moderate SNR values, signifying the usefulness of the obtained analytical results. We also provide insights on the conditions at which the proposed splitting receiver has significant performance advantages over the traditional receivers.
\end{abstract}

\begin{IEEEkeywords}
 Splitting receiver, wireless receiver, coherent detection, envelope detection.
\end{IEEEkeywords}

\section{Introduction}
\lettrine{R}{eceiver} design is an essential element in the evolution of wireless communication systems. Coherent receivers are the most commonly adopted design in cellular communications and local area networks. Such receivers perform coherent detection (CD) by extracting both the in-phase and quadrature components of the input signal. Recently, non-coherent receivers have been extensively utilized in the internet of things (IoT) and future communication systems due to its potential for enabling low-cost, low-power and straightforward implementation~\cite{gao2019energy, ma2021channel}.

Roughly speaking, there are two types of non-coherent receivers used in communication systems: envelope detection (ED) receiver and power detection (PD) receiver.
ED is the most commonly used non-coherent detector in radio-frequency (RF) communications~\cite{cheng2017an, el2020a}.
The widely seen ED circuits in practice use either a half-wave or full-wave rectifier followed by a low-pass filter, the output signal of which captures the envelope of the input RF signal~\cite{lyons2017digital}.
The PD receiver is more commonly adopted in optical communications~\cite{fa2009channel, el2020adptive, la2009on}.
{Note that while PD can still be used in RF communications, it commonly requires active components. In contrast, ED circuits are usually passive and hence have much lower cost and energy consumption.}
%In the free-space optical communication, the information is modulated through the optical intensity or power~\cite{el2020adptive}. The PD receiver uses a  active photodetector to measure the incident optical intensity of the incoming light and produces an output signal proportional to the detected power~\cite{la2009on}. { In addition, the ED circuit is commonly passive and has a much lower complexity than an active PD circuit and thus has a much lower cost and energy consumption.}

Different from the traditional coherent and non-coherent receiver designs mentioned above, a fundamentally new receiver architecture was recently proposed in \cite{Liu2017A}, \cite{wang2020on}, named as splitting receiver. This receiver architecture jointly uses both coherent and non-coherent processing for signal detection. In the original design of~\cite{wang2020on}, PD is adopted for non-coherent processing, resulting in the PD-CD splitting receiver.
The work in~ \cite{wang2020on} showed that the PD-CD splitting receiver can achieve substantial performance improvement over the traditional receivers. This was an exciting result demonstrating the huge potential of the splitting receiver architecture.
{In the context of RF wireless communication systems, ED is a much more suitable detection method  and has a much lower complexity than PD. This motivates us to propose the ED-CD splitting receiver, which gives a more practical implementation option of splitting receiver for RF communications, and investigate if the new splitting receiver can still provide performance gain compared to conventional non-splitting receivers.}

In this letter, we propose the new ED-CD splitting receiver and establish its performance limit through analyzing the achievable mutual information. Specifically, we conduct a high signal-to-noise ratio (SNR) analysis and derive an approximation of the achievable mutual information. This approximation also enables us to analytically obtain the optimal design of the splitting ratio which is the key design parameter of the splitting receiver. Our numerical results verify the accuracy of the analytical approximation even at moderately low SNR. This also implies that the optimal splitting ratio analytically obtained at high SNR can directly be used at moderate SNR for achieving best data rate performance. The numerical results also identify the conditions at which the ED-CD splitting receiver gives significant performance advantages over the traditional receivers that only use either ED or CD for signal detection.

\section{The Proposed ED-CD Splitting receiver}
\begin{figure}[t]
\centering
\includegraphics[width=0.85 \linewidth]{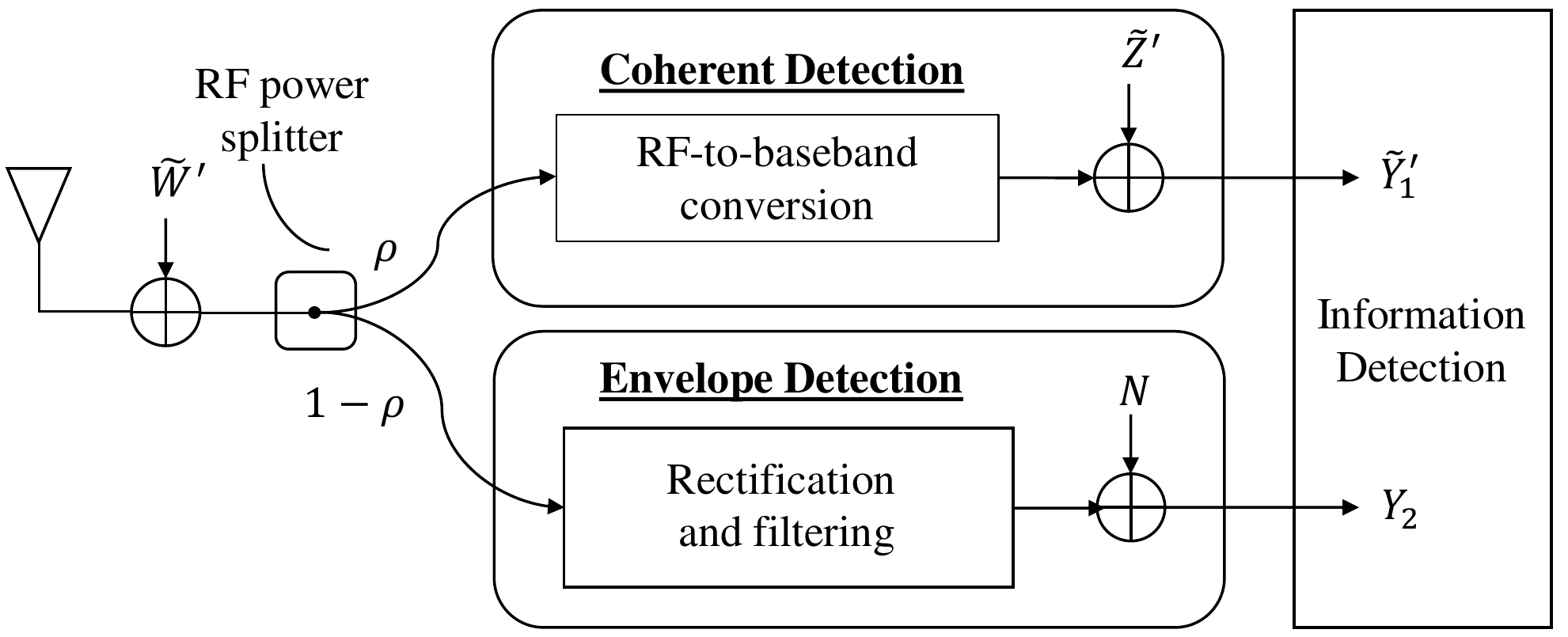}
\caption{The proposed ED-CD splitting receiver.}
\label{fig:Fig1sysmod}
%\vspace{-0.35cm}
\end{figure}

The proposed ED-CD splitting receiver design is presented in \figref{fig:Fig1sysmod}.
The RF transmitted signal propagates through a channel with coefficient $\tilde{h}$.  In this work, we assume that $\tilde{h}$ is perfectly known by the receiver. The received RF signal is divided into two streams with a power splitting ratio $\rho$, where $\rho \in [0,1]$. One stream goes through the CD processing, while the other stream goes through the ED processing.

The baseband digitized signals after the CD and ED processing are respectively given by
%${{{\tilde{Y}}}^{'}_1} = \sqrt{\rho}\,(\sqrt{P}\tilde{h}{{\tilde{X}}}+{{\tilde{W}^{'}}})+{{\tilde{Z}^{'}}}$ and ${{{Y}}_2} = \sqrt{1-\rho}\,\big|\sqrt{P}\tilde{h}{{\tilde{X}}}+{{\tilde{W}^{'}}}\big|+{{N}}$,
\begin{equation}
\label{equ:CDreceiverc}
{{{\tilde{Y}}}^{'}_1} = \sqrt{\rho}\,(\sqrt{P}\tilde{h}{{\tilde{X}}}+{{\tilde{W}^{'}}})+{{\tilde{Z}^{'}}},
\end{equation}
\begin{equation}
\label{equ:EDreceiverc}
{{{Y}}_2} = \sqrt{1-\rho}\,\big|\sqrt{P}\tilde{h}{{\tilde{X}}}+{{\tilde{W}^{'}}}\big|+{{N}},
\end{equation}
where ${{\tilde{X}}}$ is the transmitted signal with normalized variance, $P$ is the average transmit power of the signal, ${\tilde{W}^{'}}$ denotes the antenna noise, and the processing noises introduced in the CD and ED streams are denoted by $\tilde{Z}^{'}$ and ${N}$ respectively. Note that ${{\tilde{Y}}}^{'}_1$, $\tilde{h}$, $\tilde{X}$, ${\tilde{W}^{'}}$ and $\tilde{Z}^{'}$ are complex-valued, while $Y_2$ and $N$ are real-valued.

%=\sqrt{1-\rho}\big|e^{j\phi}\big|\big|\sqrt{P}{|\tilde{h}|}{{\tilde{X}}}+{{\tilde{W}}}\big|+{{N}}

{ For the convenience of the later analysis, we rewrite the channel coefficient as  $\tilde{h}\triangleq|\tilde{h}|e^{j\phi}$ and ${{{Y}}_2}\triangleq \sqrt{1-\rho}\big|\sqrt{P}{|\tilde{h}|e^{j\phi}}{{\tilde{X}}}+e^{j\phi}{{\tilde{W}}}\big|+{{N}}$, and define ${{{\tilde{Y}}}_1} \triangleq e^{-j\phi}{{{\tilde{Y}}}^{'}_1}$, ${\tilde{W}}\triangleq e^{-j\phi}{{\tilde{W}}^{'}}$ and $\tilde{Z}\triangleq e^{-j\phi}{{\tilde{Z}^{'}}}$.} The equivalent baseband signals are given by
\begin{equation}
\label{equ:CDreceiver}
{{{\tilde{Y}}}_1} = \sqrt{\rho}\,(\sqrt{P}|\tilde{h}|{{\tilde{X}}}+{{\tilde{W}}})+{{\tilde{Z}}},
\end{equation}
\begin{equation}
\label{equ:EDreceiver}
{{{Y}}_2} = \sqrt{1-\rho}\,\big|\sqrt{P}{|\tilde{h}|}{{\tilde{X}}}+{{\tilde{W}}}\big|+{{N}}.
\end{equation}

For ease of reference, we call the variables ${\tilde{W}}$, $\tilde{Z}$ and ${{N}}$ as the equivalent antenna noise, (CD) conversion noise, and (ED) rectifier noise, respectively, and they follow zero-mean Gaussian distribution with variances  $\sigma_{\textrm{A}}^2$, $\sigma_{\textrm{cov}}^2$ and $\sigma_{\textrm{rec}}^2$, respectively. Again, note that ${{{\tilde{Y}}}_1}$, ${\tilde{W}}$ and $\tilde{Z}$ are complex-valued.

{ In the practical communication systems, the conversion noise power $\sigma_{\textrm{cov}}^2$  is commonly much greater than the rectifier noise power $\sigma_{\textrm{rec}}^2$~\cite{Hen2001,flicker}. For example, the value of $\sigma_{\textrm{rec}}^2$ is -85 dBm  is -85 dBm~\cite{flicker}, while $\sigma_{\textrm{cov}}^2$ is -70~dBm~\cite{zhou2013wireless, mat1999digital}. Apart from the relatively strength of the ED and CD processing noises, we should also put them in comparison with the antenna noise. From~\cite{zhou2013wireless}, a typical antenna noise  power is approximately $-104$ dBm. Of course this antenna noise power accounts for only the thermal noise, while the antenna noise in our system model should also include any potential interference signals received which can be much stronger than thermal noise.}

Compared with the previous PD-CD splitting receiver proposed in~\cite{wang2020on}, the difference in the ED-CD splitting receiver lies in \eqref{equ:EDreceiver} due to the ED processing. Although this may seem to be a small difference in terms of the mathematical model, it does represent a different receiver design in practice. Also, as we will discuss at the end of the next section, there are some fundamental differences in the parameter design and performance between the previous PD-CD splitting receiver and the new ED-CD splitting receiver.

\section{Mutual Information Performance Analysis}

To characterize the fundamental performance limit of the proposed receiver, we aim to derive the mutual information achievable based on the signal model in \eqref{equ:CDreceiver} and \eqref{equ:EDreceiver}. Note that ${{{\tilde{Y}}}_1}$ and ${{{Y}}_2}$ are jointed used for making the detection. Therefore, the mutual information between the input ${{\tilde{X}}}$  and output $({{{\tilde{Y}}}_1}, {{{Y}}_2})$ is expressed as
\begin{align}
\label{equ:MISPreceiver}
&\mathcal{I}\left({{\tilde{X}}};{{{\tilde{Y}}}_1},{{{Y}}_2}\right) =\mathcal{H}\left({{{\tilde{Y}}}_1},{{{Y}}_2}\right)-\mathcal{H}\left({{{\tilde{Y}}}_1},{{{Y}}_2}\big|{{\tilde{X}}}\right)\notag\\
&=-\int_{{{{Y}}_2}}\int_{{{{\tilde{Y}}}_1}}f_{{{{\tilde{Y}}}_1},{{{Y}}_2}}({{{\tilde{y}}}_1},{{{y}}_2})\log_2\left(f_{{{{\tilde{Y}}}_1},{{{Y}}_2}}({{{\tilde{y}}}_1},{{{y}}_2})\right) \,\mathrm{d}{{{\tilde{y}}}_1}\mathrm{d}{{{y}}_2}\notag\\
&\,\,\,\,\,\,\,\,+\int_{{{\tilde{X}}}}\int_{{{\tilde{Y}_1}}}\int_{{{{Y_2}}}}f_{{{\tilde{X}}}}(\tilde{x})f_{{{\tilde{Y}}_1},{{{Y_2}}}}(\tilde{y}_1,{y_2}|\tilde{x})\notag\\
&\,\,\,\,\,\,\,\,\,\,\,\,\,\,\,\log_2 \left(f_{{{\tilde{Y}_1}},{{{Y_2}}}}(\tilde{y}_1,{y_2}|\tilde{x})\right)\,\mathrm{d}{y_2}\mathrm{d}\tilde{y}_1\mathrm{d}\tilde{x},
\end{align}
where the joint probability density
function (PDF) of $({{{\tilde{Y}}}_1},{{{Y}}_2})$ is given by
\begin{align}
\label{equ:pdfy1y2}
&f_{{{{\tilde{Y}}}_1},{{{Y}}_2}}({{{\tilde{y}}}_1},{{{y}}_2})\notag\\
&=\int_{\tilde{X}}\int_{\tilde{W}}f_{{{{\tilde{Y}}}_1},{{{Y}}_2}}({{{\tilde{y}}}_1},{{{y}}_2}|\tilde{x},\tilde{w})
f_{{{\tilde{X}}}}(\tilde{x})f_{{{\tilde{W}}}}(\tilde{w})\,\mathrm{d}{\tilde{w}}\mathrm{d}{\tilde{x}}\notag\\
&=\!\!\int_{\tilde{X}}\int_{\tilde{W}}\!\!f_{{{{\tilde{Y}}}_1}}({{{\tilde{y}}}_1}|\tilde{x},\tilde{w})f_{{{{Y}}_2}}({{{y}}_2}|\tilde{x},\tilde{w})
f_{{{\tilde{X}}}}(\tilde{x})f_{{{\tilde{W}}}}(\tilde{w})\,\mathrm{d}{\tilde{w}}\mathrm{d}{\tilde{x}},
\end{align}
and the conditional joint PDF $f_{{{\tilde{Y}_1}},{{{Y_2}}}}(\tilde{y}_1,{y_2}|\tilde{x})$ is
$\int_{\tilde{W}}\!\!f_{{{{\tilde{Y}}}_1}}({{{\tilde{y}}}_1}|\tilde{x},\tilde{w})f_{{{{Y}}_2}}({{{y}}_2}|\tilde{x},\tilde{w})f_{{{\tilde{W}}}}(\tilde{w})\,\mathrm{d}{\tilde{w}}$,
where $f_{{{\tilde{X}}}}(\tilde{x})$ is the PDF of the normalized input signal, $f_{{{\tilde{W}}}}(\tilde{w})$ follows the complex Gaussian distribution, which is expressed as $\mathcal{CN}\big(0,\sigma_{\textrm{A}}^2\big)$. The conditional PDFs $f_{{{{\tilde{Y}}}_1}}({{{\tilde{y}}}_1}|\tilde{x},\tilde{w})$ and $f_{{{{Y}}_2}}({{{y}}_2}|\tilde{x},\tilde{w})$ are $\mathcal{CN}\big(\sqrt{\rho}(\sqrt{P}|\tilde{h}|\tilde{x}+\tilde{w}),\sigma_{\textrm{cov}}^2\big)$ and $\mathcal{N}\,\big( \sqrt{1-\rho}\big|\sqrt{P}{|\tilde{h}|}{{\tilde{x}}}+{{\tilde{w}}}\big|,\sigma_{\textrm{rec}}^2\big)$, respectively.

We observe that the mutual information expression in \eqnref{equ:MISPreceiver} needs evaluation of a large number of integrals and the complexity is extremely high.
In addition, the mutual information depends on the distribution of $\tilde{X}$. As finding the optimal input distribution is an extremely challenging task, in this work, we only consider a widely adopted approach of assuming Gaussian input signal, i.e., $\tilde{X}\sim \mathcal{CN}(0,1)$. Note that Gaussian input is optimal for the traditional CD receivers. In what follows, we aim to give an accurate approximation of the mutual information in the high SNR regime.

\begin{figure}[t]
\subfigure[$\sigma_{\textrm{{A}}}^2=\sigma_{\textrm{{cov}}}^2=1,\sigma_{\textrm{{rec}}}^2=0.01$.]{
%\begin{minipage}[t]{0.5\linewidth}
\includegraphics[width=1.65in]{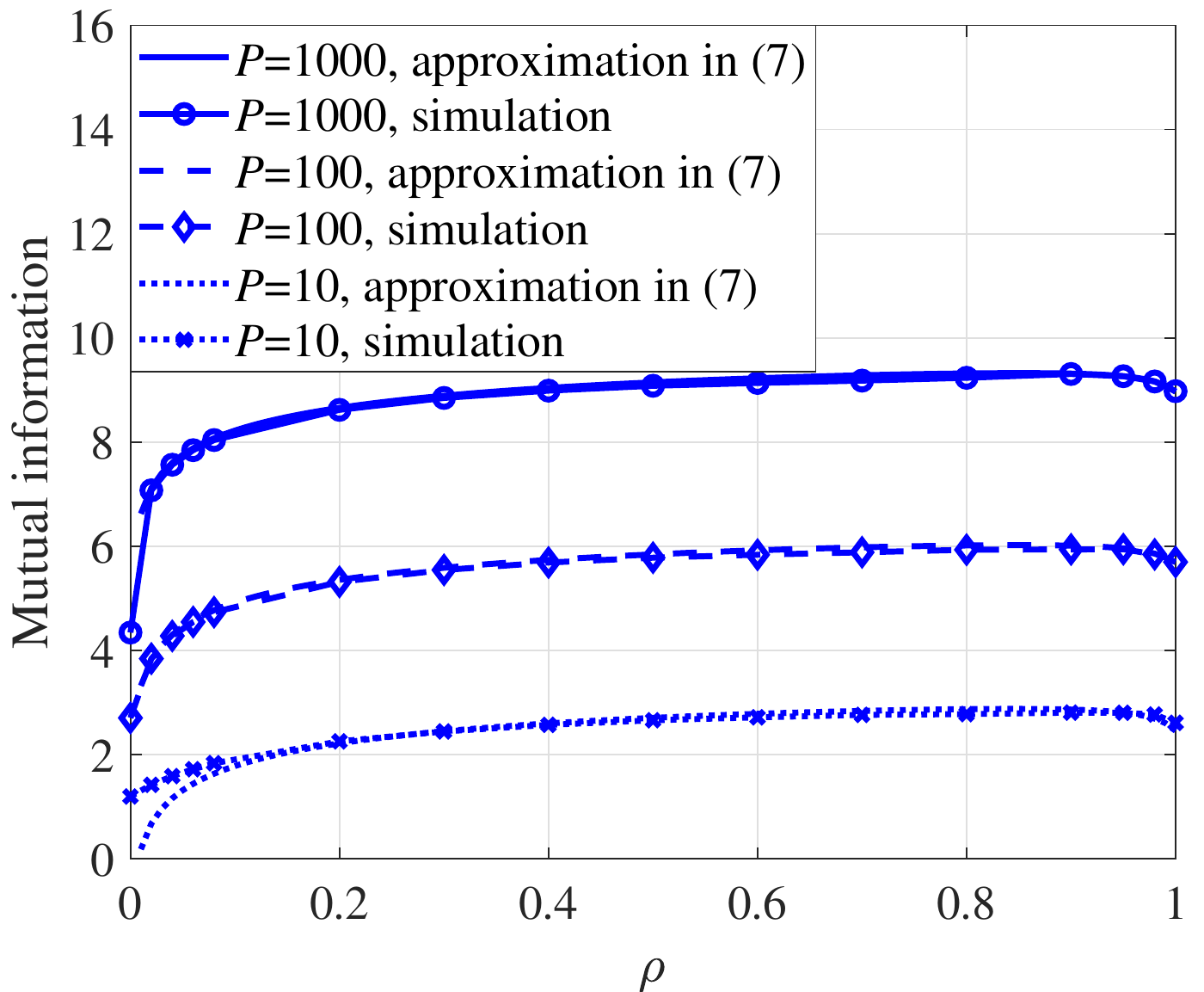}
%\end{minipage}%
}%
\subfigure[$\sigma_{\textrm{{A}}}^2=\sigma_{\textrm{{rec}}}^2=0.01, \sigma_{\textrm{{cov}}}^2=1$.]{
%\begin{minipage}[t]{0.5\linewidth}
\includegraphics[width=1.69in]{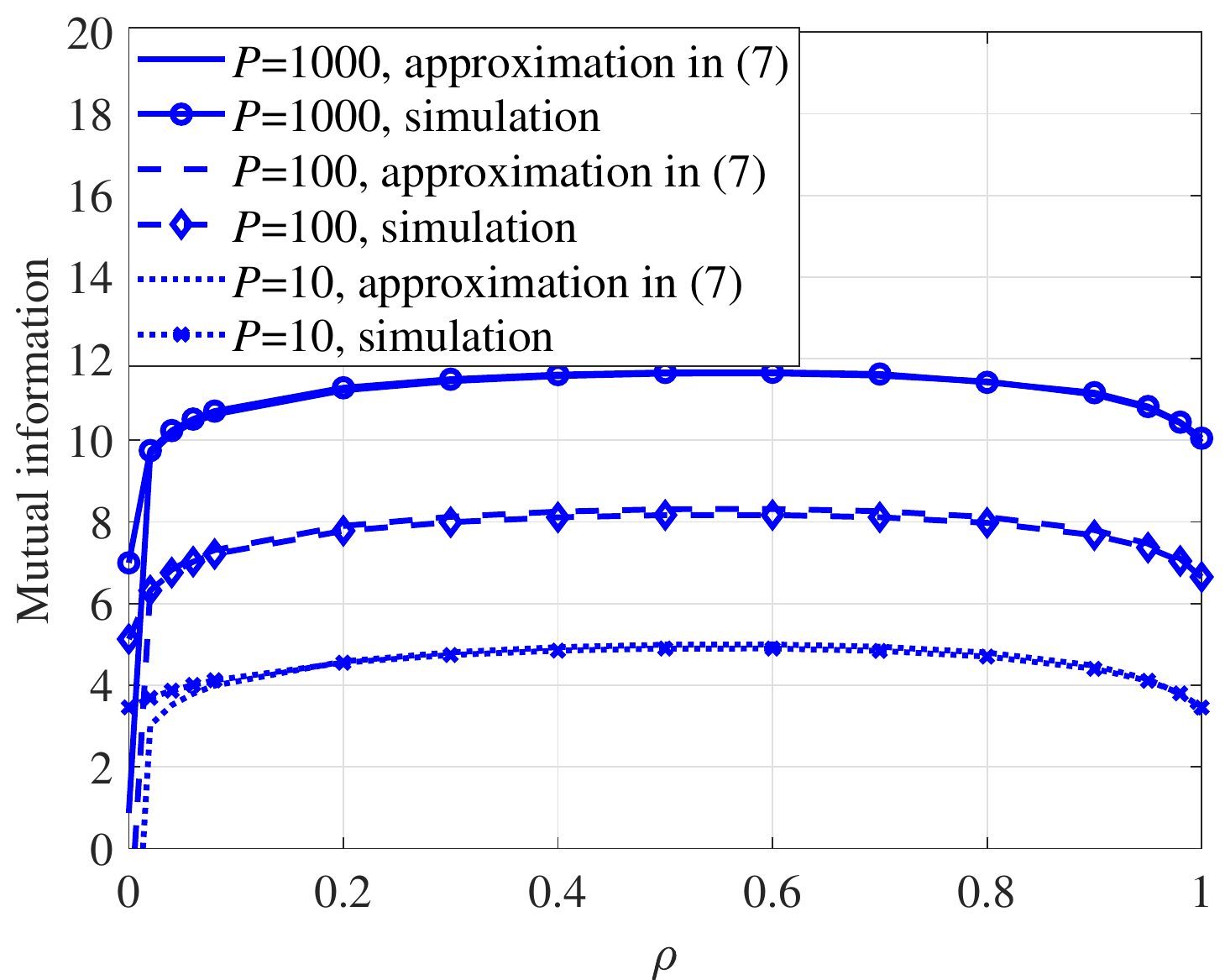}
%\end{minipage}%
}%
\caption{Mutual information versus the power splitting ratio $\rho$.}
\label{fig:MI2}
%\vspace{-0.35cm}
\end{figure}

\begin{prop}
In high SNR, the achievable mutual information of the splitting receiver with $\rho\in (0,1]$ can be approximated as
%\begin{equation}
\begin{align}
\label{equ:MIhighSNR}
&\mathcal{I}({{\tilde{X}}};{{{\tilde{Y}}}_1},{{{Y}}_2})\approx \frac{1}{2}\log_2\left(\frac{\rho(P|\tilde{h}|^2+\sigma_{\textrm{{A}}}^2)^2}{P|\tilde{h}|^2(\rho\sigma_{\textrm{{A}}}^2+\sigma_{\textrm{{cov}}}^2)}\right)+\notag\\
&\frac{1}{2}\log_2\biggl(\frac{\sigma_{\textrm{{cov}}}^2(1-\rho)(P|\tilde{h}|^2+\sigma_{\textrm{{A}}}^2)+2\rho\sigma_{\textrm{{rec}}}^2P|\tilde{h}|^2}{2\sigma_{\textrm{{rec}}}^2\sigma_{\textrm{A}}^2\rho+\sigma_{\textrm{{cov}}}^2\sigma_{\textrm{A}}^2(1-\rho)+2\sigma_{\textrm{{rec}}}^2\sigma_{\textrm{cov}}^2}\biggr).
\end{align}
Importantly, this approximation is asymptotically tight as $P$ approaches infinity.
The optimal splitting ratio $\rho^*$ in the high SNR regime is
\begin{align}
\label{equ:oprho1-2}
\rho^*=
\begin{cases}
\Upsilon, & \sigma_{\textrm{{cov}}}^2> 4 \sigma_{\textrm{{rec}}}^2,\\
1, & \textrm{else},
 \end{cases}
\end{align}
where
\begin{align}
\label{equ:oprho1-3-1} \Upsilon=\frac{\sigma_{\textrm{{cov}}}^2(\sigma_{\textrm{{cov}}}^2-2\sigma_{\textrm{{rec}}}^2)(\sigma_{\textrm{{A}}}^2+2\sigma_{\textrm{{rec}}}^2)-\sqrt{2\Psi}}{\sigma_{\textrm{{A}}}^2(\sigma_{\textrm{{cov}}}^2-2 \sigma_{\textrm{{rec}}}^2)(\sigma_{\textrm{{cov}}}^2-4\sigma_{\textrm{{rec}}}^2)},
\end{align}
and
$\Psi\!=\!\sigma_{\textrm{{cov}}}^4(\sigma_{\textrm{{cov}}}^2-2 \sigma_{\textrm{{rec}}}^2)(\sigma_{\textrm{{A}}}^2+\sigma_{\textrm{{cov}}}^2-2\sigma_{\textrm{{rec}}}^2)\sigma_{\textrm{{rec}}}^2(\sigma_{\textrm{{A}}}^2+2\sigma_{\textrm{{rec}}}^2)$.
\emph{Proof}: See Appendix A.\hfill
\label{proposition:prop3}
\end{prop}

{ Note that the result in Proposition 1 is derived under the assumption of Gaussian input distribution, while it is not clear what is the optimal input for the ED-CD splitting receiver. Therefore, finding the optimal input distribution for the ED-CD splitting receiver is a very challenging open problem.}

%\begin{coro}
%The minimum value of $\Upsilon$ is close to 0.5 but not equal to 1 when $\frac{\sigma_{\textrm{{cov}}}^2}{\sigma_{\textrm{{rec}}}^2}\rightarrow \infty$ and $\frac{\sigma_{\textrm{{A}}}^2}{\sigma_{\textrm{{rec}}}^2}\rightarrow 0$ .
%
%\emph{Proof}: See Appendix B.\hfill
%\label{corollary:coro1}
%\end{coro}
\begin{table*}[h]
	\centering
	\caption{Optimal Power Splitting Ratio and Mutual Information Performance Gain}
\begin{tabular}{|m{2.3cm}<{\centering}|m{0.4cm}<{\centering}|m{0.4cm}<{\centering}|m{0.7cm}<{\centering}|m{0.4cm}<{\centering}|m{0.4cm}<{\centering}|m{0.7cm}<{\centering}|m{0.4cm}<{\centering}|m{0.4cm}<{\centering}|m{0.7cm}<{\centering}|m{0.4cm}<{\centering}|m{0.4cm}<{\centering}|m{0.7cm}<{\centering}|m{0.6cm}<{\centering}|m{0.4cm}<{\centering}|m{0.4cm}<{\centering}|}
%\begin{tabular}{|p{2.3cm}|l|l|l|l|l|l|l|l|l|l|l|l|l|l|l|}
		\hline %\multirow{2}{*}
		{\multirow{2}{*}{Noise Conditions}} & \multicolumn{3}{c|}{$P=10$} & \multicolumn{3}{c|}{$P=10^2$} & \multicolumn{3}{c|}{$P=10^3$} & \multicolumn{3}{c|}{$P=10^4$} & \multicolumn{3}{c|}{$P\rightarrow\infty$ (analytical)} \\ \cline{2-16}
		& \multicolumn{1}{c|}{$\rho^*$}             & \multicolumn{1}{c|}{$G_{\textrm{MI}}$}             & \multicolumn{1}{c|}{$G_{\textrm{MI}}\%$ }          & \multicolumn{1}{c|}{$\rho^*$ }              &\multicolumn{1}{c|}{$G_{\textrm{MI}}$}                &\multicolumn{1}{c|} {$G_{\textrm{MI}}\%$ }            & \multicolumn{1}{c|}{$\rho^*$ }              & \multicolumn{1}{c|}{$G_{\textrm{MI}}$}               & \multicolumn{1}{c|}{$G_{\textrm{MI}}\%$}             & \multicolumn{1}{c|}{$\rho^*$ }             & \multicolumn{1}{c|}{$G_{\textrm{MI}}$}               & \multicolumn{1}{c|}{$G_{\textrm{MI}}\%$}              &\multicolumn{1}{c|}{ $\rho^*$ }               &\multicolumn{1}{c|} {$G_{\textrm{MI}}$}              &\multicolumn{1}{c|} {$G_{\textrm{MI}}\%$ }            \\ \hline\hline

\multicolumn{1}{|c|}{\begin{tabular}[c]{@{}c@{}}$\sigma_{\textrm{A}}^2=0.01$\\ $\sigma_{\textrm{cov}}^2=1$\\ $\sigma_{\textrm{rec}}^2=1$\end{tabular}} &  1             &  0            &   $0$           & 1              &    0          &  $0 $           &  1             &   0         &  $0 $           & 1              & 0              &  $0$            &   $1$            &  0            &   \multicolumn{1}{c|}{0}             \\ \hline

\multicolumn{1}{|c|}{\begin{tabular}[c]{@{}c@{}}$\sigma_{\textrm{A}}^2=0.01$\\ $\sigma_{\textrm{cov}}^2=1$\\ $\sigma_{\textrm{rec}}^2=0.1$\end{tabular}}
& 0.63& 0.18& $5.1\%$&0.63 & 0.30&$4.5\%$ &0.63 & 0.31&$3.1\%$ &0.63 &0.31 &$2.4\%$ &0.63 &0.31& \multicolumn{1}{c|}{0} \\ \hline

\multicolumn{1}{|c|}{\begin{tabular}[c]{@{}c@{}}$\sigma_{\textrm{A}}^2=0.01$\\ $\sigma_{\textrm{cov}}^2=1$\\ $\sigma_{\textrm{rec}}^2=0.01$\end{tabular}}
& 0.56& 1.56& $45.2\%$&0.56 & 1.68&$25.3\%$ &0.56 & 1.68&$17.0\%$ &0.56 &1.68 &$12.8\%$ &0.56 &1.69& \multicolumn{1}{c|}{0} \\ \hline		

\multicolumn{1}{|c|}{\begin{tabular}[c]{@{}c@{}}$\sigma_{\textrm{A}}^2=0.01$\\ $\sigma_{\textrm{cov}}^2=1$\\ $\sigma_{\textrm{rec}}^2=0.001$\end{tabular}}
& 0.71& 2.57& $74.6\%$&0.71 & 2.69&$40.5\%$ &0.71 & 2.71&$27.2\%$ &0.71 &2.71 &$20.4\%$ &0.71 &2.71& \multicolumn{1}{c|}{0} \\ \hline		

\multicolumn{1}{|c|}{\begin{tabular}[c]{@{}c@{}}$\sigma_{\textrm{A}}^2=1$\\ $\sigma_{\textrm{cov}}^2=1$\\ $\sigma_{\textrm{rec}}^2=1$\end{tabular}}
&1 & 0&$0$ &1 & 0&$0$ &1 &0 &$0$ & 1& 0& $0$&1 &0& \multicolumn{1}{c|}{0} \\ \hline	

\multicolumn{1}{|c|}{\begin{tabular}[c]{@{}c@{}}$\sigma_{\textrm{A}}^2=1$\\ $\sigma_{\textrm{cov}}^2=1$\\ $\sigma_{\textrm{rec}}^2=0.1$\end{tabular}}
& 0.74& 0.01& $0.6\%$&0.77 & 0.09&$1.5\%$ &0.77 & 0.10&$1.1\%$ &0.77 &0.10 &$0.8\%$ &0.77 &0.10& \multicolumn{1}{c|}{0} \\ \hline

\multicolumn{1}{|c|}{\begin{tabular}[c]{@{}c@{}}$\sigma_{\textrm{A}}^2=1$\\ $\sigma_{\textrm{cov}}^2=1$\\ $\sigma_{\textrm{rec}}^2=0.01$\end{tabular}}
& 0.84& 0.29& $11.4\%$&0.85 & 0.35&$6.2\%$ &0.85 & 0.36&$4.0\%$ &0.85 &0.36 &$2.9\%$ &0.85 &0.36& \multicolumn{1}{c|}{0} \\ \hline

\multicolumn{1}{|c|}{\begin{tabular}[c]{@{}c@{}}$\sigma_{\textrm{A}}^2=1$\\ $\sigma_{\textrm{cov}}^2=1$\\ $\sigma_{\textrm{rec}}^2=0.001$\end{tabular}}
& 0.94& 0.40& $15.3\%$&0.94 & 0.45&$7.9\%$ &0.94 & 0.45&$5.1\%$ &0.94 &0.45 &$3.7\%$ &0.94 &0.45& \multicolumn{1}{c|}{0} \\ \hline		
	\end{tabular}
	\label{tab:MIan}
\end{table*}
To verify the accuracy of the approximation in \eqnref{equ:MIhighSNR}, \figref{fig:MI2}~(a) and \figref{fig:MI2}~(b) depict the  mutual information approximation and the simulated mutual information against the power splitting ratio with different signal power. We observe that the analytical approximation in \eqnref{equ:MIhighSNR} matches with the simulation for all different system parameter settings, with the only exception when the splitting ratio is very small.
In addition, with an appropriate choice of the splitting ratio, the ED-CD splitting receiver achieves a higher mutual information as compared to the traditional ED receiver ($\rho = 0$) and CD receiver ($\rho = 1$). The optimal splitting ratio found in \eqref{equ:oprho1-2} is accurate in all observed curves in \figref{fig:MI2}. Therefore, the numerical results in \figref{fig:MI2}  have validated the accuracy of the analytical results in Proposition~\ref{proposition:prop3} for different system parameter settings including moderate SNR values.

%{ In the practical communication systems, the conversion noise power is much greater than the rectifier noise power. This justifies the assumption that $\sigma_{\textrm{cov}}^2\gg\sigma_{\textrm{rec}}^2$.}
%and Corollary~\ref{corollary:coro1}.

%\section{Mutual Information Performance Gain}
In order to further investigate how much performance gain the ED-CD splitting receiver brings in as compared to  the traditional ED and CD receivers, we adopt the metrics defined in ~\cite{wang2020on}, namely the mutual information performance gain $G_{\textrm{MI}}$ and its relative performance gain $G_{\textrm{MI}}\%$. Specifically, $G_{\textrm{MI}}$ is the difference between the mutual information achieved by the splitting receiver and that achieved by the best of the traditional receivers, while $G_{\textrm{MI}}\%$ is the percentage difference.
If $G_{\textrm{MI}}>0$, the splitting receiver achieves higher mutual information compared to the CD and ED receivers.
\begin{prop}
In high SNR  and with the optimal splitting ratio, the mutual information performance gain of the ED-CD splitting receiver is given by
\begin{equation}
\label{equ:MIgain2}
\begin{aligned}
\mathop{\lim} \limits_{P\rightarrow\infty}G_{\textrm{MI}}=
\begin{cases}
 \beta, &  \sigma_{\textrm{{cov}}}^2> 4 \sigma_{\textrm{{rec}}}^2,\\
  0, &  \textrm{else},
\end{cases}
 \end{aligned}
\end{equation}
\begin{equation}
\label{equ:MIgain3no-1}
\begin{aligned}
\mathop{\lim} \limits_{P\rightarrow\infty}G_{\textrm{MI}}\%= 0\%,
\end{aligned}
\end{equation}
where $\beta=\log_2 \left(\Upsilon\right)+\frac{1}{2}\log_2 \left(\frac{K}{M}\right)>0$. The variables $M$ and $K$ are determined by the noise variances and are defined in Appendix B.

\emph{Proof}: See Appendix B.\hfill  Note that between the traditional ED receiver and CD receiver, it is intuitive to understand (and proven in the Appendix for high SNR) that the CD receiver achieves a higher mutual information because it makes use of both magnitude and phase information of the received signal. Hence, the mutual information performance gain is the difference between the mutual information achieved by the splitting receiver and that achieved by the CD receiver.
\label{proposition:prop4}
\end{prop}

In order to draw insights on the performance improvement under different SNRs and different relative noise strengths, we provide numerical results in Table~\ref{tab:MIan}, which shows the results of the ED-CD splitting receivers in terms of the optimal power splitting ratio $\rho^*$, mutual information gain $G_{\textrm{MI}}$ and its percentage gain $G_{\textrm{MI}}\%$ for different values of the noise variances and transmit power $P$.

We first discuss the trend in the optimal splitting ratio $\rho^*$ as the transmit power $P$ increases. We observe that the value of $\rho^*$ in each row almost does not change as the transmit power increases. That is, the optimal splitting ratio  $\rho^*$ of the ED-CD splitting receiver is insensitive to the signal power $P$ and is only related to the noise variances, which is consistent with the conclusions obtained from Proposition~\ref{proposition:prop3}.
It implies that the design of optimal splitting ratio obtained analytically at high SNR also works very well and does not need to be changed for a wide range of moderate SNRs.

Next we discuss the mutual information gain. We find that there is no performance gain for the ED-CD splitting receiver when the conversion noise variance $\sigma_{\textrm{{cov}}}^2$ is equal to the rectifier noise variance $ \sigma_{\textrm{{rec}}}^2$ as shown in the first and fifth rows of~Table~\ref{tab:MIan}. When the rectifier noise variance $ \sigma_{\textrm{{rec}}}^2$ is much smaller than the conversion noise variance $\sigma_{\textrm{{cov}}}^2$, the ED-CD splitting receiver can provide significant performance gain. These observations are in agreement with Proposition~\ref{proposition:prop3}, which says that the optimal splitting ratio is less than 1 when the rectifier noise is much smaller than the conversion noise (more specifically 4 times smaller). Clearly when the optimal splitting ratio is 1, the splitting receiver becomes the same as the CD receiver, hence, there is no performance gain. In addition, we observe that as the antenna noise variance $\sigma_{\textrm{{A}}}^2$ decreases, the performance gain increases.
Therefore, one can conclude that the performance gain of the ED-CD splitting receiver is most substantial when the rectifier noise is much smaller than the conversion noise and the antenna noise is relatively small as compared to these processing noises. For example, in the third row when $\sigma_{\textrm{{A}}}^2=0.01$, $\sigma_{\textrm{{cov}}}^2=1$, and $\sigma_{\textrm{{rec}}}^2=0.01$, the ED-CD splitting receiver provides $45.2\%$ higher mutual information at $P = 10$ and $25.3\%$ higher mutual information at $P = 100$, compared to the traditional CD receiver.
These favorable noise conditions can be found in practical scenarios: In an interference-free scenario, the antenna noise is almost at the thermal noise level, which is much smaller than the processing noises.
Moreover, the rectifier noise variance $\sigma_{\textrm{{rec}}}^2$ is commonly much lower than the conversion noise variance $\sigma_{\textrm{{cov}}}^2$ in the practical circuits~\cite{Hen2001,flicker}. Hence, the ED-CD splitting receiver can provide significant performance improvement in practical wireless communication systems.

\begin{remark}(\emph{Comparison between the proposed ED-CD splitting receiver and the original PD-CD splitting receiver in ~\cite{wang2020on}}):
 In terms of similarity, both splitting receiver designs can achieve higher mutual information than using either the traditional coherent receiver or traditional non-coherent receiver. And this performance gain is most substantial at low to moderate operating SNRs and when the receiver noise is dominated by the processing noises rather than the antenna noise. On the other hand, there are some major differences: (i) The optimal splitting ratio of the proposed ED-CD receiver stays at a constant value for a wide range of the operating SNRs and can be much smaller than 1 even at high SNRs, while the optimal splitting ratio of the PD-CD receiver always increases towards 1 as the operating SNR increases. (ii) {The performance gain of the ED-CD receiver depends heavily on the relative strength of the rectifier noise in the ED circuit and the conversion noise in the CD circuit, while the performance gain of the PD-CD receiver has much less dependence on the strength of the active rectifier noise in the PD circuit}. For these two aspects, the high SNR asymptotic behaviours of the two different splitting receivers are fundamentally different. Therefore, many of the design and performance insights previously obtained on the PD-CD splitting receiver are not applicable to the newly proposed ED-CD splitting receiver in this work.

 { In addition, the practical advantage of the ED-CD splitting receiver over the PD-CD splitting receiver is that the passive ED circuit has a much lower complexity than the active PD circuit and thus has a much lower cost and energy consumption~\cite{Na2009}. This advantage of the ED-based solution is more significant when the receiver antennas scale up leading to the increasing of detection circuits. The low complexity ED circuit introduces smaller processing noise than the active PD. A detailed quantitative performance comparison between the two different splitting receivers requires detailed circuit-level modelling and noise analysis and is an interesting problem for future work.}

\end{remark}
\section{Conclusions}
In this letter, we have presented a new ED-CD splitting receiver, which jointly uses the envelope detection and coherent detection. Compared to the original PD-CD splitting receiver, the newly proposed ED-CD splitting receiver is more practically applicable in RF communications. Our analysis has provided accurate closed-form approximation to understand the mutual information performance of the proposed receiver. The optimal splitting ratio has also been derived at high SNR and shown to be accurate for a wide range of operating conditions. Numerical results have revealed that the proposed splitting receiver can achieve significant performance gain over the traditional receivers when the processing noise in the ED circuit is much smaller than the processing noise in the CD circuit and the antenna noise is relatively small as compared to these processing noises.

\begin{appendices}
\section{ Proof of Proposition 1}
To analyze the mutual information of the ED-CD splitting receiver, we utilize the property of mutual information invariance under a change of coordinates. Following the similar steps of the analysis of the PD-CD receiver in  ~\cite[Appendix A]{wang2020on}, we firstly represent the Cartesian coordinate based random variables as the cone-normal coordinate based random variables, and then analyze the approximated mutual information in high SNR.

Different from the paraboloid-normal coordinate system adopted in~\cite{wang2020on}, the cone of the cone-normal coordination system of this letter is given below due to the received signal expressions \eqref{equ:CDreceiver} and \eqref{equ:EDreceiver}
\setcounter{equation}{0}
\renewcommand\theequation{A.\arabic{equation}}
\begin{equation}
\label{equ:CN}
c_M=k\sqrt{\frac{1-\rho}{\rho}}\sqrt{c_I^2+c_Q^2},
\end{equation}
where $k\triangleq\frac{\sigma_{\textrm{cov}}}{\sqrt{2}\sigma_{\textrm{rec}}}$; $c_I$, $c_Q$, and $c_M$ are the three axes of Cartesian coordinate system of the in-phase-quadrature-magnitude (I-Q-M) space.

Based on the cone-normal coordinate system in ~\eqnref{equ:CN} and
following the same steps in ~\cite[Appendix A]{wang2020on}, the asymptotic mutual information can be obtained as
\begin{align}
\label{equ:MIend}
\mathcal{I}({{\tilde{X}}};{{{\tilde{Y}}}_1}, {{{Y}}_2})
&=\log_2\left(\Theta_1\zeta^2\sqrt{\frac{k^2\Theta_2\zeta^2}{\Theta_1}+1}\right)\notag\\
& -\frac{1}{2}\log_2\left( \frac{\Theta_1\sigma_{\textrm{A}}^2+\sigma_{\textrm{cov}}^2}{(\Theta_1+\Theta_2k^2)\sigma_{\textrm{A}}^2+\sigma_{\textrm{cov}}^2}\right),
\end{align}
where $\zeta^2=1+\frac{\sigma_{\textrm{A}}^2}{P|\tilde{h}|^2}$, $k^2=\frac{\sigma_{\textrm{cov}}^2}{2\sigma_{\textrm{rec}}^2}$, $\Theta_1\triangleq \rho $ and $\Theta_2\triangleq 1-\rho $. \eqnref{equ:MIhighSNR} can be obtained after simplifications.

The mutual information approximation in \eqnref{equ:MIhighSNR} can be rewritten as $\frac{1}{2}\log_2(f(\rho))$. Knowing that the function $\log(\cdot) $ is monotonic, we only need to find the optimal splitting ratio $\rho^*$ that maximizes $f(\rho)$.
In order to compute $\rho^{*}$, we first calculate derivative function $f(\rho)$ as
 \begin{align}
\label{equ:MIhighSNRsim22}
\!\!\!\!f'\!(\rho)\!\!\triangleq \!\!\frac{A\rho^2 +B\rho +C}{\!(\!\rho \sigma_{\textrm{{A}}}^2\!\!+\!\sigma_{\textrm{{cov}}}^2\!)^2 \!\big(\!(\rho\!-\!\!1\!) \sigma_{\textrm{{A}}}^2 \sigma_{\textrm{{cov}}}^2\!\!-\!\!2\rho \sigma_{\textrm{{A}}}^2 \sigma_{\textrm{{rec}}}^2\!\!-\!\!2 \sigma_{\textrm{{cov}}}^2 \sigma_{\textrm{{rec}}}^2\!\big)^2},
 \end{align}
 where
 \begin{align}
\label{equ:MIhighSNRsim22-1}
 &A=P \sigma_{\textrm{{A}}}^2 \sigma_{\textrm{{cov}}}^6-6 P \sigma_{\textrm{{A}}}^2 \sigma_{\textrm{{cov}}}^4 \sigma_{\textrm{{rec}}}^2+8 P \sigma_{\textrm{{A}}}^2 \sigma_{\textrm{{cov}}}^2 \sigma_{\textrm{{rec}}}^4\notag \\
 &\,\,\,\,\,\,\,\,\,-2 \sigma_{\textrm{{A}}}^6 \sigma_{\textrm{{cov}}}^2 \sigma_{\textrm{{rec}}}^2+\sigma_{\textrm{{A}}}^4 \sigma_{\textrm{{cov}}}^6-4 \sigma_{\textrm{{A}}}^4 \sigma_{\textrm{{cov}}}^4 \sigma_{\textrm{{rec}}}^2,\notag\\
 &B=-2 P \sigma_{\textrm{{A}}}^2 \sigma_{\textrm{{cov}}}^6+4 P \sigma_{\textrm{{A}}}^2 \sigma_{\textrm{{cov}}}^4 \sigma_{\textrm{{rec}}}^2-4 P \sigma_{\textrm{{cov}}}^6 \sigma_{\textrm{{rec}}}^2\notag\\
 &\,\,\,\,\,\,\,\,\,+8 P \sigma_{\textrm{{cov}}}^4 \sigma_{\textrm{{rec}}}^4-2 \sigma_{\textrm{{A}}}^4 \sigma_{\textrm{{cov}}}^6-4 \sigma_{\textrm{{A}}}^2 \sigma_{\textrm{{cov}}}^6 \sigma_{\textrm{{rec}}}^2,\notag \\
 &C=P \sigma_{\textrm{{A}}}^2 \sigma_{\textrm{{cov}}}^6+2 P \sigma_{\textrm{{cov}}}^6 \sigma_{\textrm{{rec}}}^2+\sigma_{\textrm{{A}}}^4 \sigma_{\textrm{{cov}}}^6+2 \sigma_{\textrm{{A}}}^2 \sigma_{\textrm{{cov}}}^6 \sigma_{\textrm{{rec}}}^2.
 \end{align}

Then, making the numerator of  $f'(\rho)=0$, the root that maximizes $f(\rho)$ is  $\rho^{*}$.
Letting $P\rightarrow\infty$ and keeping the dominate terms, the numerator of \eqnref{equ:MIhighSNRsim22} is approximated as
 \begin{align}
\label{equ:MIhighSNRsim3}
&f'_{\textrm{num}}(\rho)\triangleq(P  \sigma_{\textrm{{A}}}^2 \sigma_{\textrm{{cov}}}^6-6 P  \sigma_{\textrm{{A}}}^2 \sigma_{\textrm{{cov}}}^4 \sigma_{\textrm{{rec}}}^2+8 P \sigma_{\textrm{{A}}}^2\sigma_{\textrm{{cov}}}^2 \sigma_{\textrm{{rec}}}^4)\rho^2+ \notag\\
&(4 P \sigma_{\textrm{{A}}}^2\sigma_{\textrm{{cov}}}^4 \sigma_{\textrm{{rec}}}^2-2 P \sigma_{\textrm{{A}}}^2 \sigma_{\textrm{{cov}}}^6-4 P\sigma_{\textrm{{cov}}}^6 \sigma_{\textrm{{rec}}}^2+
8 P \sigma_{\textrm{{cov}}}^4 \sigma_{\textrm{{rec}}}^4)\rho\notag\\
&+P \sigma_{\textrm{{A}}}^2 \sigma_{\textrm{{cov}}}^6+ 2P \sigma_{\textrm{{cov}}}^6 \sigma_{\textrm{{rec}}}^2.
\end{align}

When $\sigma_{\textrm{{A}}}^2\neq 0$ and $\sigma_{\textrm{{cov}}}^4-6 \sigma_{\textrm{{cov}}}^2 \sigma_{\textrm{{rec}}}^2+8 \sigma_{\textrm{{rec}}}^4=(\sigma_{\textrm{{cov}}}^2-2 \sigma_{\textrm{{rec}}}^2)(\sigma_{\textrm{{cov}}}^2-4\sigma_{\textrm{{rec}}}^2)= 0$, we have the equality
$f'_{\textrm{num}}(\rho)\triangleq 2( 2\sigma_{\textrm{{rec}}}^2-\sigma_{\textrm{{cov}}}^2)( \sigma_{\textrm{{A}}}^2 +2\sigma_{\textrm{{rec}}}^2)\rho+ \sigma_{\textrm{{cov}}}^2(\sigma_{\textrm{{A}}}^2 + 2 \sigma_{\textrm{{rec}}}^2) =0$, which has the single root as $
\rho^*=1$.
When $\sigma_{\textrm{{A}}}^2\neq 0$ and $(\sigma_{\textrm{{cov}}}^2-2 \sigma_{\textrm{{rec}}}^2)(\sigma_{\textrm{{cov}}}^2-4\sigma_{\textrm{{rec}}}^2)\neq 0$, the function $f'_{\textrm{num}}(\rho)=0$ has two roots:
\begin{align}
\label{equ:oprho1-3A}
 \Upsilon=\frac{\sigma_{\textrm{{cov}}}^2(\sigma_{\textrm{{cov}}}^2-2\sigma_{\textrm{{rec}}}^2)(\sigma_{\textrm{{A}}}^2+2\sigma_{\textrm{{rec}}}^2)-\sqrt{2\Psi}}{\sigma_{\textrm{{A}}}^2(\sigma_{\textrm{{cov}}}^2-2 \sigma_{\textrm{{rec}}}^2)(\sigma_{\textrm{{cov}}}^2-4\sigma_{\textrm{{rec}}}^2)},
\end{align}
and
\begin{align}
\label{equ:oprho1-3P}
 \Phi=\frac{\sigma_{\textrm{{cov}}}^2(\sigma_{\textrm{{cov}}}^2-2\sigma_{\textrm{{rec}}}^2)(\sigma_{\textrm{{A}}}^2+2\sigma_{\textrm{{rec}}}^2)+\sqrt{2\Psi}}{\sigma_{\textrm{{A}}}^2(\sigma_{\textrm{{cov}}}^2-2 \sigma_{\textrm{{rec}}}^2)(\sigma_{\textrm{{cov}}}^2-4\sigma_{\textrm{{rec}}}^2)},
\end{align}
where
$\Psi\!=\!\sigma_{\textrm{{cov}}}^4(\sigma_{\textrm{{cov}}}^2\!\!-\!2 \sigma_{\textrm{{rec}}}^2)(\sigma_{\textrm{{A}}}^2\!\!+\!\sigma_{\textrm{{cov}}}^2\!\!-\!2\sigma_{\textrm{{rec}}}^2)\sigma_{\textrm{{rec}}}^2(\sigma_{\textrm{{A}}}^2\!\!+\!2\sigma_{\textrm{{rec}}}^2)$.

From \eqnref{equ:oprho1-3A} and \eqnref{equ:oprho1-3P}, if $\sigma_{\textrm{{cov}}}^2<2 \sigma_{\textrm{{rec}}}^2$, the numerators and denominators  of the $\Upsilon$ and  $\Phi$ are smaller and larger than zero, respectively. Then we can obtain that $\Upsilon<\Phi<0$, and easy to prove that $f(\rho)$ is an increasing function in $0<\rho<1$, so $\rho^*=1$.
When $2 \sigma_{\textrm{{rec}}}^2<\sigma_{\textrm{{cov}}}^2<4 \sigma_{\textrm{{rec}}}^2$, the roots $\Phi<0$,  $\Upsilon>0$ and the value of the root $\Phi$ is smaller than $\Upsilon$, and it can be verified that $f(\rho)$ is an increasing function in $0<\rho<1$ and $\rho^*=1$.
When $\sigma_{\textrm{{cov}}}^2>4 \sigma_{\textrm{{rec}}}^2$, we have $\Phi >\Upsilon>0$. It can be proved that $\Upsilon$ maximizes $f(\rho)$ and $0<\Upsilon<1$, Then the optimal splitting ratio $\rho^*=\Upsilon$.
\section{Proof of Proposition 2}
According to the definition of the $G_{\textrm{MI}}$ shown in~\cite{wang2020on}, for analyzing the mutual information performance gain, we need to obtain the asymptotic mutual information of the ED receiver ($\rho=0$), CD receiver ($\rho=1$) and the ED-CD splitting receiver. The mutual information of the ED-CD splitting receiver has been achieved as shown in Proposition~\ref{proposition:prop3}. The mutual information of the CD receiver is~\cite{wang2020on}
\begin{align}
\mathcal{I}({{\tilde{X}}};{{{\tilde{Y}}}_1}, {{{Y}}_2}\!)|_{\rho=1}
=\log_2\left(1+\frac{P|\tilde{h}|^2}{\sigma_{\textrm{A}}^2+\sigma_{\textrm{cov}}^2}\right).
\end{align}
An upper bound on the mutual information of the ED receiver is given by~\cite{zhou2013wireless, katz2004capacity}
\begin{align}
\label{equ:NOcapupp}
\mathcal{I}({{\tilde{X}}};{{{\tilde{Y}}}_1}, {{{Y}}_2}\!)|_{\rho=0}&=\mathcal{I}({{\tilde{X}}}; {{{Y}}_2}\!)\notag \\
&\leq  \frac{1}{2}\!\log_2\!\!\left(\frac{P|\tilde{h}|^2}{\sigma_{\textrm{A}}^2}\right)\!\!+\!\!\frac{1}{2}\!\!\left(\!\log_2\frac{e}{2\pi}\!\right).
\end{align}

Based on \eqnref{equ:NOcapupp}, the upper bound on mutual information of the ED receiver scales with $P$ as $\frac{1}{2}\log_2 P$. While the achievable mutual information of the CD receiver scales with $P$ as $\log_2 P$. Therefore,  the CD receiver achieves much higher mutual information than the ED receiver as $P\rightarrow\infty$, i.e.,
$\mathcal{I}({{\tilde{X}}};{{{\tilde{Y}}}_1},{{{Y}}_2})|_{\rho=0}\ll \mathcal{I}({{\tilde{X}}};{{{\tilde{Y}}}_1},{{{Y}}_2})|_{\rho=1}$.
Thus, we have~\cite{wang2020on}
\begin{align}
\label{equ:MIgain1}
G_{\textrm{MI}}\!= \!\sup\{\mathcal{I}({{\tilde{X}}};&{{{\tilde{Y}}}_1},{{{Y}}_2}): \rho\in (0,1]\}\!-\!\mathcal{I}({{\tilde{X}}};{{{\tilde{Y}}}_1},{{{Y}}_2})|_{\rho=1}.
\end{align}

Then, the performance improvement of the splitting receiver is expressed as
\begin{align}
\label{equ:MIgain2a}
&\mathop{\lim} \limits_{P\rightarrow\infty}G_{\textrm{MI}}\!= \! \mathcal{I}(\!{{\tilde{X}}};{{{\tilde{Y}}}_1},\!{{{Y}}_2}\!)|_{\rho=\rho^*}\!-\!\mathcal{I}(\!{{\tilde{X}}};{{{\tilde{Y}}}_1},\!{{{Y}}_2}\!)|_{\rho=1}\notag\\
&=\frac{1}{2}\log_2 \!\!\left(\frac{2 P|\tilde{h}|^2 \rho^* \sigma_{\textrm{{rec}}}^2-(\rho^*-1) \sigma_{\textrm{{cov}}}^2 (P|\tilde{h}|^2+\sigma_{\textrm{{A}}}^2)}{\sigma_{\textrm{{A}}}^2 (-\rho^* \sigma_{\textrm{{cov}}}^2+2 \rho^* \sigma_{\textrm{{rec}}}^2+\sigma_{\textrm{{cov}}}^2)+2 \sigma_{\textrm{{cov}}}^2 \sigma_{\textrm{{rec}}}^2}\right)+\notag\\
&\frac{1}{2}\!\log_2\!\! \left(\frac{(P|\tilde{h}|^2+\sigma_{\textrm{{A}}}^2)^2\rho^{*}(\sigma_{\textrm{{A}}}^2+\sigma_{\textrm{{cov}}}^2)^2}{P|\tilde{h}|^2 (\rho^* \sigma_{\textrm{{A}}}^2+\sigma_{\textrm{{cov}}}^2)(P|\tilde{h}|^2+\sigma_{\textrm{{A}}}^2+\sigma_{\textrm{{cov}}}^2)^2}\!\!\right).
\end{align}

Based on \eqnref{equ:MIgain2a}, when $\rho^*=1$,  $G_{\textrm{MI}}=0$.
When $\rho^*=\Upsilon$, $G_{\textrm{MI}}$ is simplified as
\begin{align}
\label{equ:MIgain2e}
\mathop{\lim} \limits_{P\rightarrow\infty}G_{\textrm{MI}}\!&= \! \mathcal{I}(\!{{\tilde{X}}};{{{\tilde{Y}}}_1},\!{{{Y}}_2}\!)|_{\rho=\Upsilon}\!-\!\mathcal{I}(\!{{\tilde{X}}};{{{\tilde{Y}}}_1},\!{{{Y}}_2}\!)|_{\rho=1}\notag\\
&=\log_2 \left(\Upsilon\right)+\frac{1}{2}\log_2 (K/M),
\end{align}
where
\begin{align}
&K=(\sigma_{\textrm{{A}}}^2+\sigma_{\textrm{{cov}}}^2)^2 (\sigma_{\textrm{{cov}}}^4-6 \sigma_{\textrm{{cov}}}^2 \sigma_{\textrm{{rec}}}^2+8 \sigma_{\textrm{{rec}}}^4)^2
(2 \sigma_{\textrm{{A}}}^2 \sigma_{\textrm{{cov}}}^2 \sigma_{\textrm{{rec}}}^2+\notag \\
&\,\,\,\,\,\,\,\,\,\,\,\,2 \sigma_{\textrm{{cov}}}^4 \sigma_{\textrm{{rec}}}^2-4 \sigma_{\textrm{{cov}}}^2 \sigma_{\textrm{{rec}}}^4-\sqrt{2\Psi}),\notag \\
&M=\!\!(\sigma_{\textrm{{A}}}^2 \sigma_{\textrm{{cov}}}^2 (\sigma_{\textrm{{cov}}}^2\!\!-\!\!2 \sigma_{\textrm{{rec}}}^2)\!\!+\!\!\sigma_{\textrm{{cov}}}^6\!\!-\!4 \sigma_{\textrm{{cov}}}^4 \sigma_{\textrm{{rec}}}^2\!\!+\!4 \sigma_{\textrm{{cov}}}^2 \sigma_{\textrm{{rec}}}^4\!\!-\!\sqrt{2\Psi} )\notag\\
&\,\,\,\,\,\,\,\,\,\,\,\,(4 \sigma_{\textrm{{A}}}^2 \sigma_{\textrm{{rec}}}^2 (\sigma_{\textrm{{cov}}}^2-2 \sigma_{\textrm{{rec}}}^2)+2 \sigma_{\textrm{{cov}}}^4 \sigma_{\textrm{{rec}}}^2-4 \sigma_{\textrm{{cov}}}^2 \sigma_{\textrm{{rec}}}^4-\sqrt{2\Psi}) \notag\\
 &\,\,\,\,\,\,\,\,\,\,\,\,\sigma_{\textrm{{cov}}}^4 (\sigma_{\textrm{{A}}}^2+2 \sigma_{\textrm{{rec}}}^2).
\end{align}
%and
%$M=\!\!(\sigma_{\textrm{{A}}}^2 \sigma_{\textrm{{cov}}}^2 (\sigma_{\textrm{{cov}}}^2\!\!-\!\!2 \sigma_{\textrm{{rec}}}^2)\!\!+\!\!\sigma_{\textrm{{cov}}}^6\!\!-\!4 \sigma_{\textrm{{cov}}}^4 \sigma_{\textrm{{rec}}}^2\!\!+\!4 \sigma_{\textrm{{cov}}}^2 \sigma_{\textrm{{rec}}}^4\!\!-\!\sqrt{2\Psi} )
%(4 \sigma_{\textrm{{A}}}^2 \sigma_{\textrm{{rec}}}^2 (\sigma_{\textrm{{cov}}}^2-2 \sigma_{\textrm{{rec}}}^2)+2 \sigma_{\textrm{{cov}}}^4 \sigma_{\textrm{{rec}}}^2-4 \sigma_{\textrm{{cov}}}^2 \sigma_{\textrm{{rec}}}^4-\sqrt{2\Psi})
% \sigma_{\textrm{{cov}}}^4 (\sigma_{\textrm{{A}}}^2+2 \sigma_{\textrm{{rec}}}^2)$.

Note that when $\rho^*=\Upsilon$,  the splitting receiver can strictly achieve higher mutual information than the CD receiver ($\rho=1$), as proved in Appendix A, i.e., $\mathcal{I}(\!{{\tilde{X}}};{{{\tilde{Y}}}_1},\!{{{Y}}_2}\!)|_{\rho^*=\Upsilon}\!>\!\mathcal{I}(\!{{\tilde{X}}};{{{\tilde{Y}}}_1},\!{{{Y}}_2}\!)|_{\rho=1}$ as $P\rightarrow \infty$. In addition, when $\rho=1$, the performance improvement gain $G_{\textrm{MI}}=0$. Therefore, when $
\rho^*=\Upsilon$, the performance improvement $G_{\textrm{MI}}>0$, i.e., $\beta >0$. Then, we can easily obtain the result related to $G_{\textrm{MI}}\%$.
\end{appendices}

\bibliographystyle{IEEEtran}
%\bibliography{ref}
%% that's all folks

\end{document}